\documentclass[pra,12pt,a4paper,showpacs,amsmath,amssymb]{revtex4}

\usepackage{color,epsfig,rotating,graphicx,subfigure}
\usepackage{amsmath,amssymb,amscd}
\usepackage{tensor}

\usepackage[latin1]{inputenc}
\usepackage[english]{babel}

\usepackage{dsfont}
\usepackage{textcomp}

\renewcommand{\Re}{{\rm Re}}
\renewcommand{\Im}{{\rm Im}}

\newcommand{\ri}{{\rm i}}
\newcommand{\re}{{\rm e}}
\newcommand{\rd}{{\rm d}}
\newcommand{\rr}{{\rm r}}
\newcommand{\rt}{{\rm t}}

\newcommand{\rs}{{\rm s}}
\newcommand{\rp}{{\rm p}}

\newcommand{\rA}{{\rm A}}

\begin{document}

%
%
\title{Statistical properties of spontaneous emission near a rough surface}

\author{S.-A. Biehs}
\altaffiliation[Present address: ]{Institut f\"{u}r Physik, Carl von Ossietzky Universit\"{a}t,
D-26111 Oldenburg, Germany.}

\author{J.-J. Greffet}

\affiliation{Laboratoire Charles Fabry, Institut d'Optique, CNRS, Universit\'{e} Paris-Sud, Campus
Polytechnique, RD128, 91127 Palaiseau cedex, France}

\date{\today}
\begin{abstract}
We study the lifetime of the excited state of an atom or molecule near a plane surface with
a given random surface roughness. In particular, we discuss the impact of the scattering of surface modes
within the rough surface. Our study is completed by considering the lateral correlation 
length of the decay rate and the variance discussing its relation to the $C_0$ correlation.
\end{abstract}

\maketitle

%
%

\section{Introduction}


The spontaneous decay rate of an excited atom or molecule is known to
depend on its environment. This effect is similar to 
Purcell's effect which has been studied theoretically and experimentally in numerous
works since the pioneering works by Purcell and Drexhage~\cite{Purcell1946,DrexhageEtAl1968}. 
In the very close proximity of a surface the decay rate increases drastically,
since the excited atom or molecule can couple to non-radiative modes. This can be related to 
the increase of the local density of states (LDOS) near a surface which
is due to evanescent modes providing more channels into which the excited atom or molecule 
can decay~\cite{FordWeber1984,Barnes1998}. 
 
Recently, this effect which allows for controlling the decay rate of atoms and molecules 
has been intensively investigated for random or disordered media. In such materials the
multiple scattering of electromagnetic modes results in the formation of speckles, i.e., 
spatial fluctuations of the LDOS. Then the spontaneous decay rate of atoms or molecules 
close or within such systems becomes a statistical quantity, which depends on the one hand on the 
local near-field environment of the source and on the other hand on the mesoscopic fluctuations 
of the random material itself. In particular, the fluorescence rate statistics or fluctuations
of the LDOS in such media has been considered 
theoretically~\cite{Shapiro1986,Shapiro1999,VanTiggelen2006,CazeEtAl2010,FroufeEtAl2007} and 
experimentally~\cite{RuijgrokEtAl2010,BirowosutoEtAl2010,SapienzaEtAl2011,Krachmalnicoff2010}.

A random rough surface is similar to a bulk disordered medium in the sense that above such a surface the LDOS shows
a spatial speckle pattern~\cite{GreffetCarminati1995}. The lifetime of an atom or molecule becomes a random quantity which depends
on the local environment of the particle and the statistical properties of the surface. Recently, the speckle pattern above random media~\cite{Apostol2003,Apostol2003b,Carminati2009} has been studied.
The goal of this work is to reconsider the impact of surface roughness on the 
spontaneous decay rate. Previous studies have considered the impact of the surface roughness
on the average decay rate for atoms or molecules near metal surfaces~\cite{Arya1982,Aravind1980,AriasEtAl1981,Xiao1989}.
 Here, we will focus on the lateral correlation of the decay rates, i.e., the correlation between the decay rates of an
atom or molecule placed at different positions above the rough surface by keeping the distance to the mean surface constant, 
and its variance. In addition, we will specifically consider the influence of surface modes for 
which the enhancement of the decay rate is very large.

For pedagogical reasons we consider a semi-infinite SiC material with a rough surface. Firstly,
we do not have to consider nonlocal effects for the considered atom-surface distances which can be quite 
important for metal surfaces~\cite{FordWeber1984,Sipe1979,PerssonLang1982,Avouris1983,BalzerEtAl1997}. Secondly, SiC has only
one well established surface resonance and is well described by a simple model~\cite{BohrenHuffman} for its permittivity. 
Nonetheless, the results we obtain are applicable to arbitrary local, homogeneous and isotropic materials. 
For the description of the surface roughness we use the perturbation theory introduced in 
Ref.~\cite{Greffet1988} up to second order within the surface profile function. 


The paper is organized as follows: In Sec.~2 we give a short introduction to the calculation
of decay rates close to a plane surface in the weak coupling limit. The effect of the surface
roughness on the mean decay rate is discussed in Sec.~3 where we also give some interpretation of
the observed roughness correction. Finally in Sec.~4 we investigate the lateral correlation and
the fluctuations of the decay rate. We finish with the conclusion in Sec.~5.

%
%

\section{Spontaneous decay rate}

\begin{figure}[Hhbt]
  \centering
  \epsfig{file=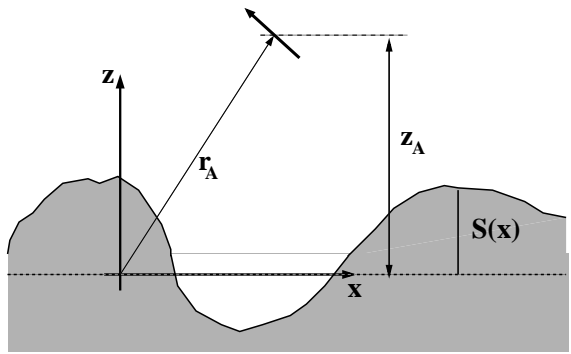, width=0.45\textwidth}
  \caption{\label{Fig:AtomRoughSurface} Sketch of a dipole at $\mathbf{r}_\rA$ in a distance $z_\rA$ above the mean of a rough surface described by a profile function $S$.}
\end{figure}

For an electric-dipole transition in the weak-coupling regime, the normalized
spontaneous decay rate of an atom or molecule placed at $\mathbf{r}_\rA$ can be expressed as~\cite{NovotnyHecht2006}
\begin{equation}
  \frac{\Gamma_i}{\Gamma_\infty} = \frac{6 \pi}{k_0} \Im \biggl[ \mathbf{e}_i^\rt \cdot \mathds{G} (\mathbf{r}_\rA,\mathbf{r}_\rA,\omega_0) \cdot \mathbf{e}_i \biggr]
\label{Eq:PurcellFactor}
\end{equation}
where $\mathbf{e}_i$ is the unit vector in the direction of the dipole transition, $\rt$ symbolizes the transposed vector, 
$k_0 = \omega_0/c$ with $\omega_0$ the frequency of 
the dipole transition. $\Gamma_\infty$ is the decay rate in free space (see for example in Ref.~\cite{NovotnyHecht2006}).
$\mathds{G}$ is the classical electric Green's dyadic for the geometry considered. In our case, this geometry
consists of a half-space of a given material characterized by its permittivity $\epsilon(\omega)$ with a 
rough surface as depicted in Fig.~(\ref{Fig:AtomRoughSurface}).

Before considering the role of surface roughness, we summarize the known results for a flat surface.
We can derive the decay rate from Eq.~(\ref{Eq:PurcellFactor}) for a dipole moment
parallel $\Gamma_\parallel$ and perpendicular $\Gamma_\perp$ to the surface by inserting the Green's dyadic from Eq.~(\ref{Eq:AppGreenFlat})
in appendix~\ref{App:GreenFlat}. We find the well-known relations~\cite{NovotnyHecht2006}
\begin{align}
    \frac{\Gamma_\parallel^{(0)}}{\Gamma_\infty} &= \frac{3}{4} \int_0^{k_0} \!\! \frac{\rd \kappa}{k_0} \, \frac{\kappa}{\gamma_\rr}\biggl\{ 1 
                                              + \Re\bigl(r_\rs \re^{2 \ri \gamma_\rr z_\rA} \bigr) 
                                           + \frac{\gamma_\rr^2}{k_0^2}\bigl[1 - \Re\bigl(r_\rp \re^{2 \ri \gamma_\rr z_\rA} \bigr)\bigr] \biggr\} \nonumber \\
                                          & \quad + \frac{3}{4} \int_{k_0}^\infty \!\! \frac{\rd \kappa}{k_0} \, \frac{\kappa}{\gamma} \re^{- 2 \gamma z_\rA} \biggl\{ \Im(r_\rs) + \Im(r_\rp) \frac{\gamma^2}{k_0^2}  \biggr\}, \label{Eq:GammaPar}\\
    \frac{\Gamma_\perp^{(0)}}{\Gamma_\infty} &= \frac{3}{2} \int_0^{k_0} \!\! \frac{\rd \kappa}{k_0} \, \frac{\kappa^3}{\gamma_\rr k_0^2}\biggl\{ 1 
                                              + \Re\bigl(r_\rp \re^{2 \ri \gamma_\rr z_\rA} \bigr) \biggr\}  
                                              + \frac{3}{2} \int_{k_0}^\infty \!\! \frac{\rd \kappa}{k_0} \, \frac{\kappa^3}{\gamma k_0^2} \re^{- 2 \gamma z_\rA} \Im(r_\rp). \label{Eq:GammaPerp}
\end{align}
Here, we have introduced the lateral wave vector $\boldsymbol{\kappa} = (k_x, k_y)^\rt$, the perpendicular wave vector $\gamma_r = \sqrt{k_0^2 - \kappa^2}$ and 
$\gamma = \sqrt{\kappa^2 - k_0^2}$; $r_\rs$ and $r_\rp$ are Fresnel's coefficient for s- and p-polarized light. Additionally, we have split
the decay rate into its radiative ($\kappa < k_0$) and non-radiative ($\kappa > k_0$) part.

\begin{figure}[Hhbt]
  \epsfig{file=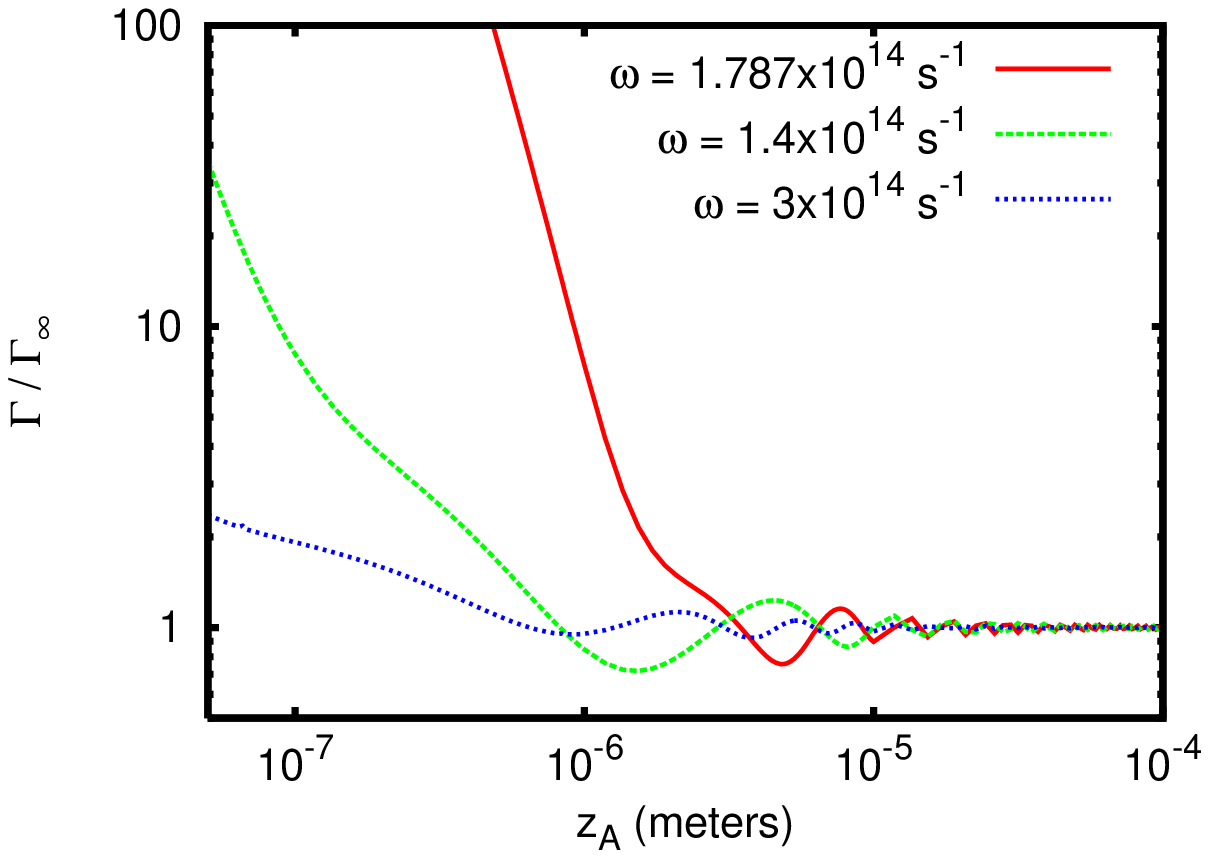, width=0.45\textwidth}
  \epsfig{file=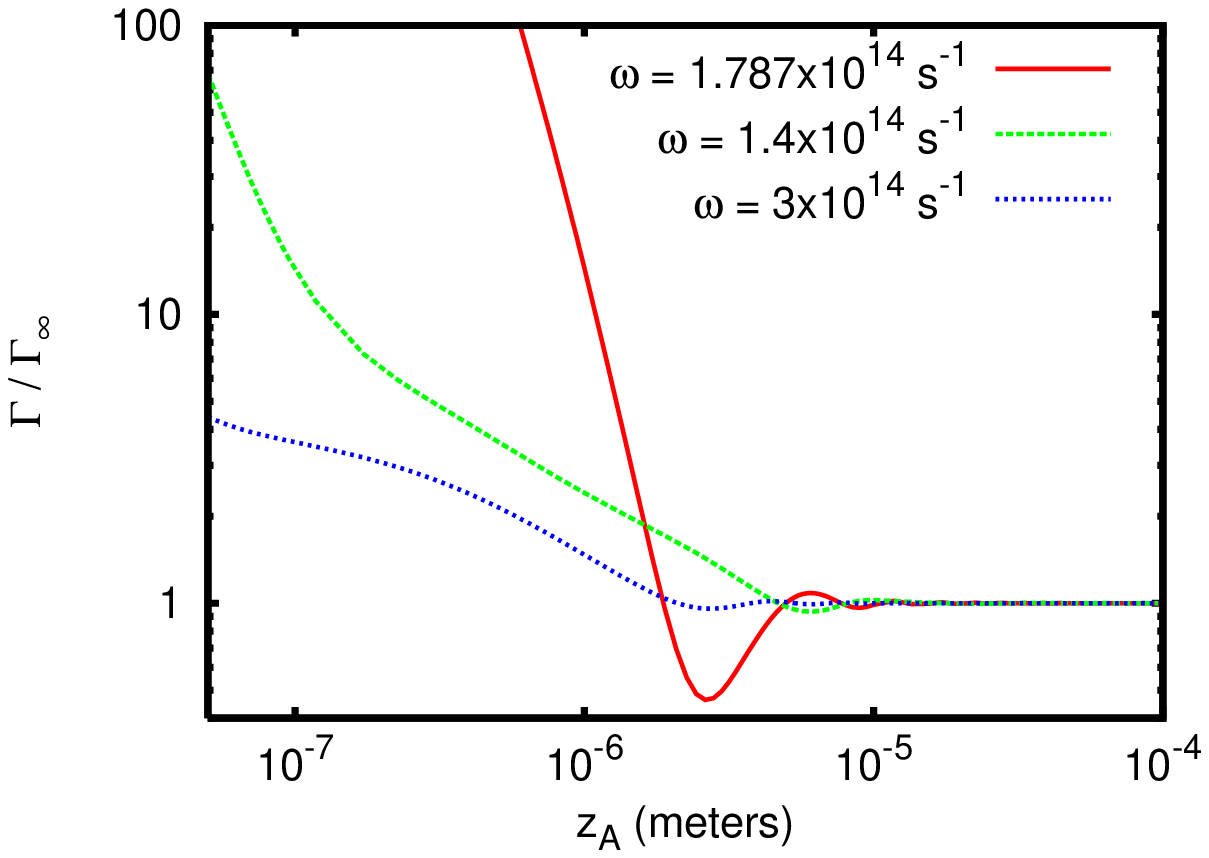, width=0.45\textwidth}
  \caption{\label{Fig:LifetimeFlatSurface} (Color online) 
           Normalized decay rate of an atom in a distance $z_\rA$ above a flat surface with its dipole moment 
           oriented (left) parallel and (right) perpendicular to the surface. 
           We have chosen the material properties of SiC at the transition frequencies
           $\omega_0 = 1.4\cdot10^{14}\,{\rm s}^{-1}, 1.787\cdot10^{14}\,{\rm s}^{-1}$ and $3\cdot10^{14}\,{\rm s}^{-1}$. }
\end{figure}

In Fig.~\ref{Fig:LifetimeFlatSurface} we show a plot of the distance dependence of the decay rates $\Gamma_\parallel$ and $\Gamma_\perp$
for an atom near a flat SiC interface. We consider a transition frequency coinciding with the SiC surface phonon resonance
at $\omega_{\omega_{\rm SPhP}} = 1.787\cdot10^{14}\,{\rm s}^{-1}$ and a transition frequency slightly smaller and slightly 
larger than $\omega_{\rm SPhP}$. In all three cases one finds the known characteristics of the decay rate near a 
flat surface~\cite{Barnes1998}: 
i) For relative large distances $z_\rA > \lambda_0 = 2 \pi c / \omega_0$ the decay rate oscillates
due to the phase change of the reflected field. 
ii) In the near-field regime with $z_\rA < \lambda_0 = 2 \pi c / \omega_0$ the decay rate is highly increased due to
the decay into nonradiative or evanescent channels. For  $\omega = 1.4\cdot10^{14}\,{\rm s}^{-1}$ and $\omega = 3\cdot10^{14}\,{\rm s}^{-1}$ the atom or molecule can decay into total internal reflection modes, whereas for $\omega = 1.787\cdot10^{14}\,{\rm s}^{-1}$
it can decay into  surface phonon polaritons. It can be expected that within this near-field regime, the decay rate
will be very sensitive to the  multiple scattering of surface waves within a rough surface for $\omega = 1.787\cdot10^{14}\,{\rm s}^{-1}$.  
iii) Finally, for a distance $z_\rA$ smaller than about $100\,{\rm nm}$ the decay rate diverges
as $z_\rA^{-3}$. This so-called quenching effect emerges from the $1/z^3$ electrostatic interaction of the atom's dipole field 
with the surface. Therefore, this effect is extremely localized so that in this extreme near-field regime
the decay rate will only be sensitive to the change of the local environment of the atom or molecule
as for example to the local change of the surface geometry due to roughness.

%
%

\section{Roughness correction to the decay rate}
Now, we turn to the effect of surface roughness on the decay rate. To this end, we
consider a stochastic surface profile function $S$ describing the deviation of the 
rough surface from flatness (see Fig.~\ref{Fig:AtomRoughSurface}). The function $S$ is modeled as a
stochastic Gaussian process with mean value and correlation function given by
\begin{align}
  \langle S(\mathbf{x}) \rangle &= 0, \\
  \langle S(\mathbf{x}) S(\mathbf{x}') \rangle &= \delta^2 \re^{-\frac{|\mathbf{x - x'}|^2}{a^2}} = \delta^2 W(|\mathbf{x - x'}|),
\end{align}
$\mathbf{x} = (x,y)^t$. The brackets $\langle \rangle$ stand for the average over an ensemble of realizations of
the surface profile $S(\mathbf{x})$; $\delta$ is the rms height
and $a$ the correlation length of the surface profile.
It follows that the Fourier components $\tilde{S}(\boldsymbol{\kappa})$ of the surface profile function
fulfill the relations
\begin{align}
  \langle \tilde{S} (\boldsymbol{\kappa}) \rangle &= 0, \\
  \langle \tilde{S} (\boldsymbol{\kappa}) \tilde{S} (\boldsymbol{\kappa}')\rangle &= (2 \pi)^2 \delta^2 
                      \delta(\boldsymbol{\kappa} + \boldsymbol{\kappa}') g(\kappa),
\end{align}
where we have introduced the surface roughness power spectrum
\begin{equation}
  g(\kappa) = \int\!\!\rd^2 x\, W(|\mathbf{x}|) \re^{- \ri \boldsymbol{\kappa} \cdot \mathbf{x}}. 
\label{Eq:PowerSpectrum}
\end{equation}
In the following calculations we will assume a gaussian correlation function $W(|\mathbf{x - x'}|) = \exp(|\mathbf{x - x'}|^2/a^2)$ so that
in this case $g(\kappa) = \pi a^2 \re^{-\frac{\kappa^2 a^2}{4}}$.
By introducing a stochastic surface profile, the fields are scattered by that surface and hence the decay rate 
becomes a stochastic processes. The reflected fields can be described by a stochastic reflection coefficient which
determines the decay rate in Eqs.~(\ref{Eq:GammaPar}) and (\ref{Eq:GammaPerp}). The statistics of the decay rate is itself determined by its mean
value and higher moments. Here, we will concentrate on the mean decay rate and in the next section we will turn to the
correlation function and the variance. By virtue of Eqs.~(\ref{Eq:GammaPar}) and (\ref{Eq:GammaPerp}) 
the mean decay rate depends on the mean reflection coefficients of the surface. The average restores the translational invariance so that the decay rate depends only on the distance to the surface $z_A$. The mean reflection coefficients
can be determined perturbatively if the surface roughness is much smaller than the wavelength $\lambda_0$.  
It has been shown in Ref.~\cite{BiehsGreffet} that by using the perturbation theory of Ref.~\cite{Greffet1988}, the correction 
to the Fresnel reflection coefficient $\Delta r_{\rs/\rp} = \langle r_{\rs/\rp} \rangle - r_{\rs/\rp}$ due to roughness is up to second-order in the surface profile given as
\begin{equation}
  \Delta r_{\rs/\rp} = - 2 \ri \gamma_\rr (D_{\rs/\rp})^2 M_{\rs/\rp} 
\end{equation}   
where
\begin{equation}
  D_\rs = \frac{\ri}{\gamma_\rr + \gamma_\rt} \quad\text{and}\quad D_\rp = \frac{\ri \epsilon}{ \gamma_\rr \epsilon + \gamma_\rt}
\end{equation}
using $\gamma_\rt = \sqrt{k_0^2 \epsilon - \kappa^2}$. The expressions for $M_\rs$ and $M_\rp$ can be found in Ref.~\cite{BiehsGreffet}.
Therefore, one can easily get the second-order correction to the decay rates 
\begin{equation}
  \Delta \Gamma_{\parallel/\perp} = \frac{\langle \Gamma_{\parallel/\perp} \rangle -  \Gamma_{\parallel/\perp}^{(0)}}{\Gamma_\infty}
\end{equation}
by replacing the reflection coefficient in Eqs.~(\ref{Eq:GammaPar})
and (\ref{Eq:GammaPerp}) by $\Delta r_{\rs}$ and $\Delta r_{\rp}$.
Now, we use the approximation for the correction to the reflection coefficient from Ref.~\cite{BiehsGreffet} 
\begin{equation}
  \Delta r_{\rs/\rp} \approx r_{\rs/\rp} 2 \kappa^2 \delta^2
\label{Eq:SDAreflect}
\end{equation}
which holds in the quasi-static regime for $\kappa \gg k_0$ and $\kappa a \gg 1$. Since this is in the quasi-static regime equivalent to distances $z_\rA \ll a$, we can conclude that for  $z_\rA \ll a$ we have 
\begin{equation}
  \Delta \Gamma_{\parallel/\perp} \approx 6 \frac{\delta^2}{z^2_\rA} \Gamma_{\parallel/\perp}^{(0)}. 
\label{Eq:SDA}
\end{equation}

\begin{figure}[Hhbt]
  \epsfig{file=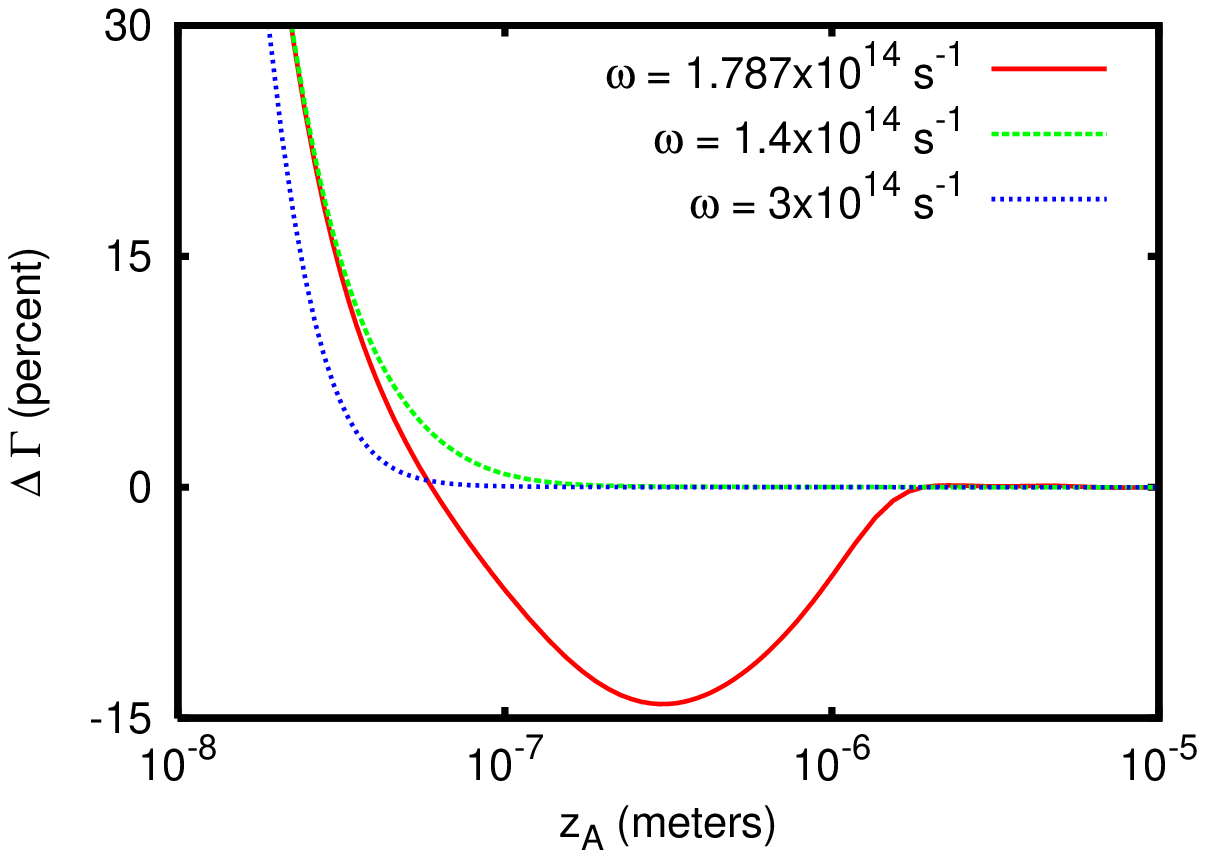, width=0.45\textwidth}
  \epsfig{file=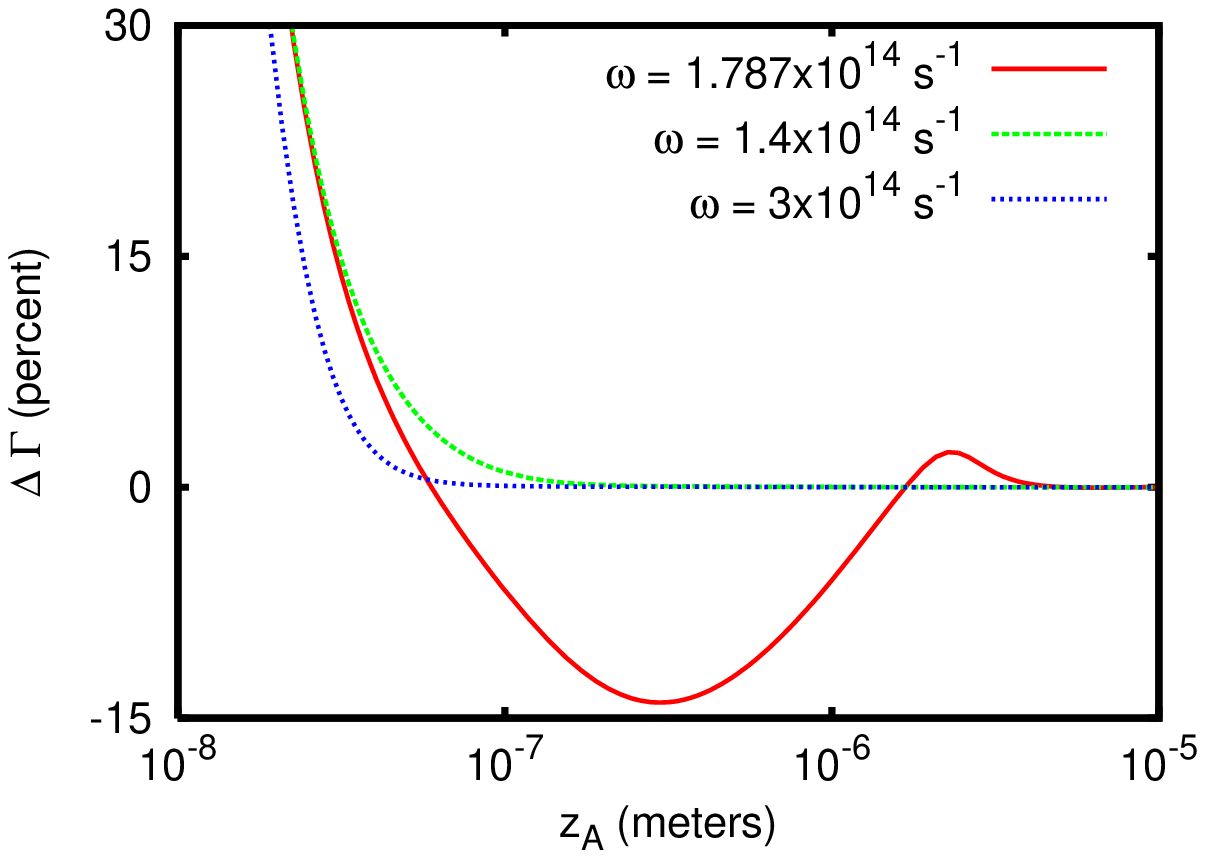, width=0.45\textwidth}
  \caption{\label{Fig:LifetimeRoughSurface} (Color online) Correction to the decay rate $\Delta \Gamma$ for an atom with its dipole moment 
           oriented (left) parallel and (right) perpendicular to the mean surface. The parameters are the same as in Fig.~\ref{Fig:LifetimeFlatSurface}. 
           For the rough surface we choose $\delta = 5\,{\rm nm}$ and $a = 200\,{\rm nm}$. }
\end{figure}

In order to illustrate the effect of roughness we plot in Fig.~\ref{Fig:LifetimeRoughSurface} the roughness correction $\Delta \Gamma_{\parallel/\perp}$ for the same frequencies as in Fig.~\ref{Fig:LifetimeFlatSurface} considering a rough surface with an rms
$\delta = 5\,{\rm nm}$ and a correlation length $a = 200\,{\rm nm}$. It can be seen that the roughness correction is 
very small in the large distance regime for $z_\rA > \lambda_0$, but can 
be relatively large for small distances, i.e., for $z_\rA < \lambda_0$ where the decay rate is very large due to the decay 
into non-radiative channels. As will be discussed in more detail in the following, electrostatic effects in the extreme near-field
for $z_\rA < 100\,{\rm nm}$ and surface phonon polaritons in the intermediate distance regime are responsible for this 
relatively large correction. The first is a local effect, whereas the latter is a multiple scattering effect.  

\begin{figure}[Hhbt]
  \centering
  \epsfig{file=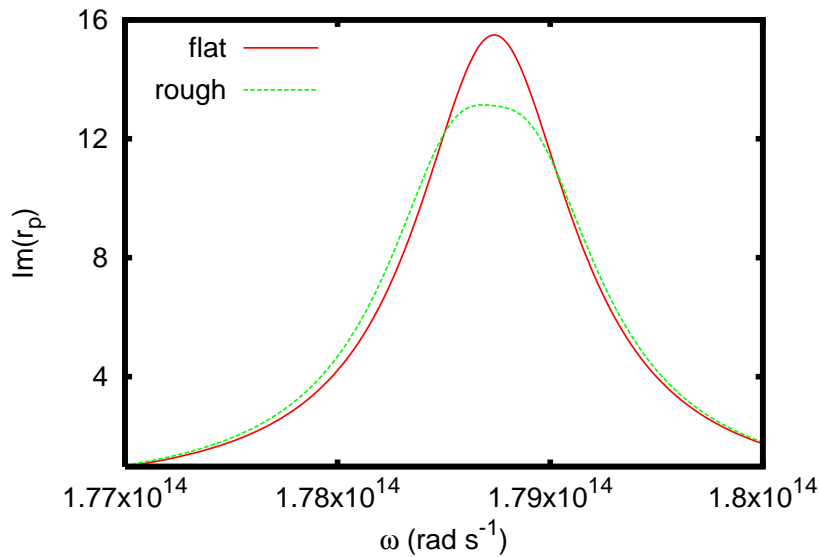, width=0.65\textwidth}
  \caption{\label{Fig:Broadening} (Color online) Plot of $\Im(r_p)$ for the flat and $\Im(\langle r_p\rangle)$ for the rough surface for $\kappa = 3.3\cdot10^{6}\,{\rm m}^{-1}$
           which corresponds approximately to a distance of $z_\rA \approx \kappa^{-1} = 3\cdot10^{-7}\,{\rm m}^{-1}$. }
\end{figure}

In the intermediate distance regime $100-1000$nm it can be seen on  Fig.~\ref{Fig:LifetimeRoughSurface}
 that the roughness correction is slightly positive at a distance of $\approx 1000$ nm when total internal reflection modes 
are excited (i.e., for $\omega_0 = 1.4\cdot10^{14}\,{\rm s}^{-1}$ and $\omega_0 = 3\cdot10^{14}\,{\rm s}^{-1}$).
For a frequency $\omega_0 = \omega_{\rm SPhP}$ surface phonon polaritons are excited in this distance regime. 
Surprisingly, the presence of roughness leads to a large negative correction indicating that the lifetime is increased. 
This effect has been studied in Ref.~\cite{BiehsGreffet} in terms of the LDOS (see Fig.~$14$ Ref.~\cite{BiehsGreffet})
and is due to the roughness induced multiple scattering of surface modes.  The scattering causes a broadening of 
the dispersion relation~\cite{Raether}. Due to this broadening as illustrated in Fig.~\ref{Fig:Broadening} the 
LDOS (which is proportional to $\Im(r_p)$) becomes smaller for 
frequencies close to $\omega_{\rm SPhP}$ for intermediate $\kappa$ or distances $z_\rA$, respectively, explaining the observed 
decrease of the decay rate. 

We now consider the very small distances $z_\rA \ll a$ regime. The roughness correction is in 
this case due to the local electrostatic interaction (quenching) of the atom dipole moment with the rough surface resulting in a positive and large roughness correction to the decay rate. This quenching effect is similar for all transition frequencies as can be seen in Fig.~\ref{Fig:LifetimeRoughSurface}. This correction can be described by the quasi-static expression in Eq.~(\ref{Eq:SDA}).
We will now see that we can retrieve this expression in Eq.~(\ref{Eq:SDA}) using a simple physical argument. If   $z_\rA \ll a$, curvature effects are negligible so that the atom feels only the local deviation of the surface from flatness.
This effect can be described by the ansatz:
\begin{equation}
  \langle \Gamma_{\parallel/\perp} \rangle \approx \langle \Gamma (d + S(\mathbf{x}))\rangle.
\end{equation}
This means that one replaces locally the surface profile by a shifted flat surface.  
Employing this approximation in Eqs.~(\ref{Eq:GammaPar}) and (\ref{Eq:GammaPerp}) we see that it is
equivalent to the replacement of the mean reflection coefficient by
\begin{equation}
  \langle r_{\rs/\rp} \rangle = r_{\rs/\rp} \langle e^{2 \ri \gamma_\rr S(\mathbf{x})} \rangle.
\end{equation} 
Note, that this is the expression for the propagating ($\kappa < k_0$) and the evanescent ($\kappa > k_0$) part.
Now, this expression can be easily evaluated, since we have assumed that $S(\mathbf{x})$ is Gaussian distributed. We find
\begin{equation}
  \langle r_{\rs/\rp} \rangle = r_{\rs/\rp} e^{- 2 \gamma_\rr^2 \delta^2}.
\end{equation}
For propagating waves such that $\kappa < k_0$, this is the well-known result of the Kirchhoff approximation~\cite{DeSanto} which holds
if $\lambda \ll a$. But for $\kappa \gg k_0$, this approximation produces the result in Eq.~(\ref{Eq:SDAreflect}) when expanding 
the exponential up to second order in the rms $\delta$. Hence, we have retrieved Eq.~(\ref{Eq:SDA}).

%
%

\section{Statistical properties of spontaneous emission}

We now examine the fluctuations and the spatial correlations of the decay rate above a random rough surface. The statistical properties of fields in random media has received a lot of attention in the past thirty years. Here, we are interested in some recent results relevant for our system. It has been shown recently that the intensity correlations above random media or materials with rough surfaces become non universal in the near field, i.e., 
they highly depend on the properties of the random media or rough surfaces~\cite{GreffetCarminati1995,EmilianiEtAl2003,Apostol2003,Apostol2003b,LaverdantEtAl2008,Carminati2010}. A remarkable connection has been established between the LDOS fluctuations and the $C_0$ correlation~\cite{Shapiro1999,VanTiggelen2006} for multiple scattering media. The $C_0$ correlation is defined as the infinite range contribution to the correlation of the intensity in 
multiple scattering media~\cite{Shapiro1999}. A simple explanation has been reported recently \cite{CazeEtAl2010}. Finally, the multiple scattering of surface modes in a random media or 
rough surface can lead to localized surface modes~\cite{BozhevolnyiEtAl1995,BozhevolnyiEtAl1996} which show a characteristic 
long tail distribution of the intensity enhancement of the fields close to the surface~\cite{BozhevolnyiCoello2001,BuilEtAl2006}. 
This effect can be neglected for surfaces with small roughnesses as shown for fractal surfaces~\cite{SanchezGil2000,SanchezGil2001}. 
Here, we focus on the lateral correlation and the variance of the decay rate above a rough surface and
we will discuss the relation of the variance to the LDOS fluctuations and $C_0$ correlation. Since the perturbative approach is 
restricted to small surface roughnesses, we will leave the problem
of localization and its relation to the distribution of decay rates for future studies. 

\subsection{Variance and correlation function}

Before evaluating the correlation function of the decay rate, we first determine the
variance which is up to second order in the surface profile given by
\begin{equation}
  \sigma^2_i = \langle \Gamma^2_i \rangle - \langle \Gamma_i \rangle^2 
             = \langle \Gamma^{(1)}_i \Gamma^{(1)}_i \rangle.
\label{Eq:Variance}
\end{equation} 
Obviously the variance is a special case of the
more general correlation function
\begin{equation}
  \langle \Gamma_i (\mathbf{r}) \Gamma_j (\mathbf{r}') \rangle = \langle \Gamma_i \Gamma_j' \rangle_{\rm spec} + \langle \Gamma_i \Gamma_j' \rangle_{\rm diff},
\end{equation}
which can be divided into a specular (depending only on the mean field) and a diffuse contribution (due to the fluctuating part of the field). 
By inserting the perturbation 
expansion $\Gamma_i \approx \Gamma_i^{(0)} +\Gamma_i^{(1)}  +\Gamma_i^{(2)} $ we find up to second order for both of these contributions
\begin{equation}
\begin{split}
   \langle \Gamma_i \Gamma_j' \rangle_{\rm spec} &= \Gamma_i^{(0)} (z)  \Gamma_j^{(0)} (z') +  \Gamma_i^{(0)} (z)  \langle \Gamma_j^{(2)} (z') \rangle \\
                                                 &\qquad  +  \langle \Gamma_i^{(2)} (z) \rangle  \Gamma_j^{(0)} (z') 
\end{split}
\label{Eq:CorrSpec}
\end{equation}
and
\begin{equation}
 \langle \Gamma_i \Gamma_j' \rangle_{\rm diff} = \langle \Gamma_i^{(1)} (\mathbf{r})  \Gamma_j^{(1)} (\mathbf{r}')  \rangle.
\end{equation}
Obviously, the specular part only depends on $z$ and $z'$. This is due to the fact that for the mean field the translational symmetry
with respect to the $x$-$y$ plane is restored after averaging, whereas for a fixed $z$ and $z'$ the diffuse part contains
lateral correlations with respect to $|\mathbf{x} - \mathbf{x}'|$. Furthermore, we note that the variance depends on the
diffuse part of the correlation function, only.

\subsection{Lateral Correlation}

Now, we focus on the lateral correlation only. Therefore, we assume $z = z' = z_\rA$. The correlation
function is then given by
\begin{equation}
  \langle \Gamma_i(\mathbf{x}) \Gamma_j(\mathbf{x}') \rangle = \langle \Gamma_i \Gamma_j \rangle_{\rm spec} + \langle \Gamma_i(\mathbf{x}) \Gamma_j(\mathbf{x}') \rangle_{\rm diff}.
\end{equation}
It is clear from Eq.~(\ref{Eq:CorrSpec}) that the specular contribution is just a constant term giving an infinite
range correlation depending on the distance $z_\rA$ only. On the other hand, the diffuse part depends on $|\mathbf{x} - \mathbf{x}'|$ and therefore contains the lateral short range correlations.  
We derive the explicit expression for the diffuse part of the correlation function 
in appendix~\ref{App:Correlation}. The result can be stated as [see Eq.~(\ref{Eq:CorrRef})]
\begin{equation}
  \frac{\bigl\langle \Gamma^{(1)}_i(\mathbf{x}) \Gamma^{(1)}_j(\mathbf{x}') \bigr\rangle}{\Gamma_\infty^2} = \frac{(3 \pi)^2}{k_0^2} 2 \Re \int\!\!\frac{\rd^2 \xi}{(2 \pi)^2} \delta^2 g(|\boldsymbol{\xi}|)
                         F_j (\boldsymbol{\xi};z_\rA) G_i (\boldsymbol{\xi};z_\rA) \re^{\ri \boldsymbol{\xi}\cdot(\mathbf{x} - \mathbf{x}')}
\label{Eq:CorrelText}
\end{equation}
where the functions $G_i$ and $F_j$ are defined in Eq. (\ref{Eq:G}) and (\ref{Eq:F}). In the following we will
discuss this expression in more detail. 

Let us focus on the evanescent regime, i.e., $z_\rA \ll \lambda$. Then   
the exponential function $\exp(\ri \gamma_\rr' z_\rA) \approx \exp(- |\boldsymbol{\kappa} \pm \boldsymbol{\xi}| z_\rA )$ 
in the integrand $a_i$ of $F_j$ and $G_i$ [see Eq.~(\ref{Eq:int_a})] acts as a low pass filter and 
restricts the contributing $\xi$ to $\xi < 1/z_\rA$. On the other hand, for a Gaussian roughness correlation 
the roughness power spectrum $g(|\xi|)$ also acts as a low
pass filter restricting the $\xi$ to $\xi < 1/a$. Therefore we can make simple approximations for Eq.~(\ref{Eq:CorrelText}) in the
two limits $a \gg z_\rA$ and $a \ll z_\rA$.

In the case $a \gg z_\rA$ the functions $F_j$ and $G_i$ can be approximated by $F_j(\mathbf{0};z_\rA)$ and $G_i(\mathbf{0};z_\rA)$. It follows 
immediately from Eq.~(\ref{Eq:CorrRef}) 
\begin{equation}
  \frac{\bigl\langle \Gamma^{(1)}_i {\Gamma^{(1)}_j}' \bigr\rangle}{\Gamma_\infty^2} \approx (3 \pi)^2 4 
                        \Im[G_j (\mathbf{0};z_\rA)] \Im[G_i (\mathbf{0};z_\rA)] \frac{\delta^2}{k_0^2} W(|\mathbf{x} - \mathbf{x}'|).
\label{Eq:CorrInText}
\end{equation}
We can conclude from this expression that the lateral correlation function reproduces the correlation function of the
surface roughness for distances such that $z_\rA \ll a$, which also holds for $z_\rA > \lambda$. 
In particular the correlation length of the lifetime correlations coincides with $a$. This means that in this distance regime
one can directly {\itshape measure the correlation of surface roughness by measuring the correlations of 
lifetimes above the surface}. In the quasi-static limit the correlation function can be further 
simplified (see appendix~\ref{App:QuasiStaticAppr} for a detailed calculation).
We find
\begin{equation}
  \frac{ \langle \Gamma_{\parallel/\perp} \Gamma_{\parallel/\perp} \rangle_{\rm diff} }{\Gamma_{\parallel/\perp}^{(0)}(z_\rA)  \Gamma_{\parallel/\perp}^{(0)}(z_\rA) } \approx
  9 \frac{\delta^2}{z^2_\rA} W(|\mathbf{x - x'}|).
\label{Eq:abiggerz}
\end{equation}

Now, in the opposite limit, where we have $a \ll z_\rA$ the roughness power spectrum $g(|\xi|)$ in Eq.~(\ref{Eq:PowerSpectrum}) can be approximated 
by $g(0) = \pi a^2$ and can be taken out of the integral. The remaining expression can be further simplified in the evanescent 
regime assuming that the most important contributions stem from $\xi \gg k_0$ and $\xi \gg k_0 |\epsilon|$. 
The resulting expression for the $\Gamma_\parallel$ can be written as (see Eq.~(\ref{Eq:CorrelPerpLarge}) appendix \ref{App:QuasiStaticAppr})
\begin{equation}
  \bigl\langle \Gamma^{(1)}_\perp (\mathbf{r}) \Gamma^{(1)}_\perp(\mathbf{r}') \bigr\rangle
                         \propto \delta^2 a^2 \frac{P_3\biggl( \frac{z_\rA}{\sqrt{z^2_\rA + (|\mathbf{x-x'}|^2}} \biggr)}{[z^2_\rA + |\mathbf{x-x'}|^2]^2},
\label{Eq:CorrelPerp}
\end{equation}
where $P_3$ is the Legendre polynomial of third power. Hence, for distances such that $a \ll z_\rA \ll \lambda$ the lateral correlation 
goes rapidly to zero for $|\mathbf{x - x'}| \gg z_\rA$. Surprisingly, the lateral correlation length only depends on $z_\rA$ and does neither
depend on the correlation length of the surface roughness nor on the properties of the material. 
For $\bigl\langle \Gamma^{(1)}_\parallel (\mathbf{x}) \Gamma^{(1)}_\parallel(\mathbf{x}') \bigr\rangle$ we find a somewhat more
complicated but similar expression in Eq.~(\ref{Eq:CorrelParLarge}) which leads to the same conclusions. We note that
a similar result was found for the intensity correlation in the near-field of a random medium~\cite{Apostol2003,Carminati2010}

\begin{figure}[Hhbt]
  \centering
  \epsfig{file = 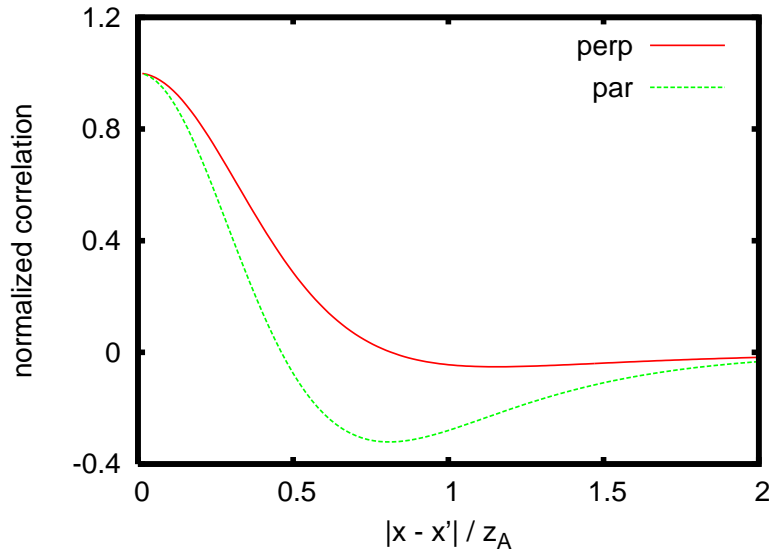, width=0.65\textwidth}
  \caption{\label{Fig:Correlation} (Color online)
    Plot of the correlation functions $\bigl\langle \Gamma^{(1)}_\perp (\mathbf{r}) \Gamma^{(1)}_\perp(\mathbf{r}') \bigr\rangle$ 
    and $\bigl\langle \Gamma^{(1)}_\parallel (\mathbf{r}) \Gamma^{(1)}_\parallel(\mathbf{r}') \bigr\rangle$ 
    in Eq.~(\ref{Eq:CorrelPerp}) and (\ref{Eq:CorrelParLarge}) over the lateral distance $(\mathbf{x} - \mathbf{x}')/z_\rA$ 
    normalized to their value at $\mathbf{x} = \mathbf{x}'$.
  }
\end{figure}

\begin{figure}[Hhbt]
  \centering
  \epsfig{file = 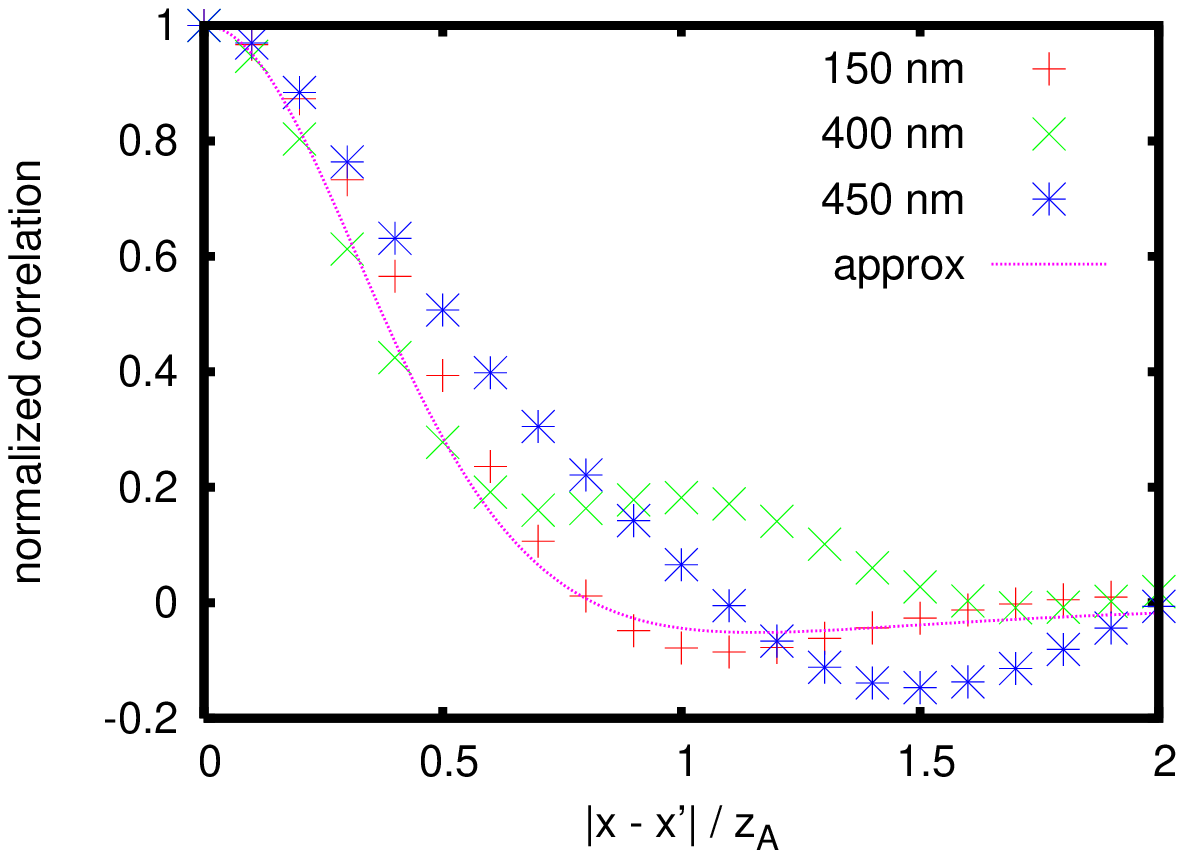, width=0.45\textwidth}
  \epsfig{file = 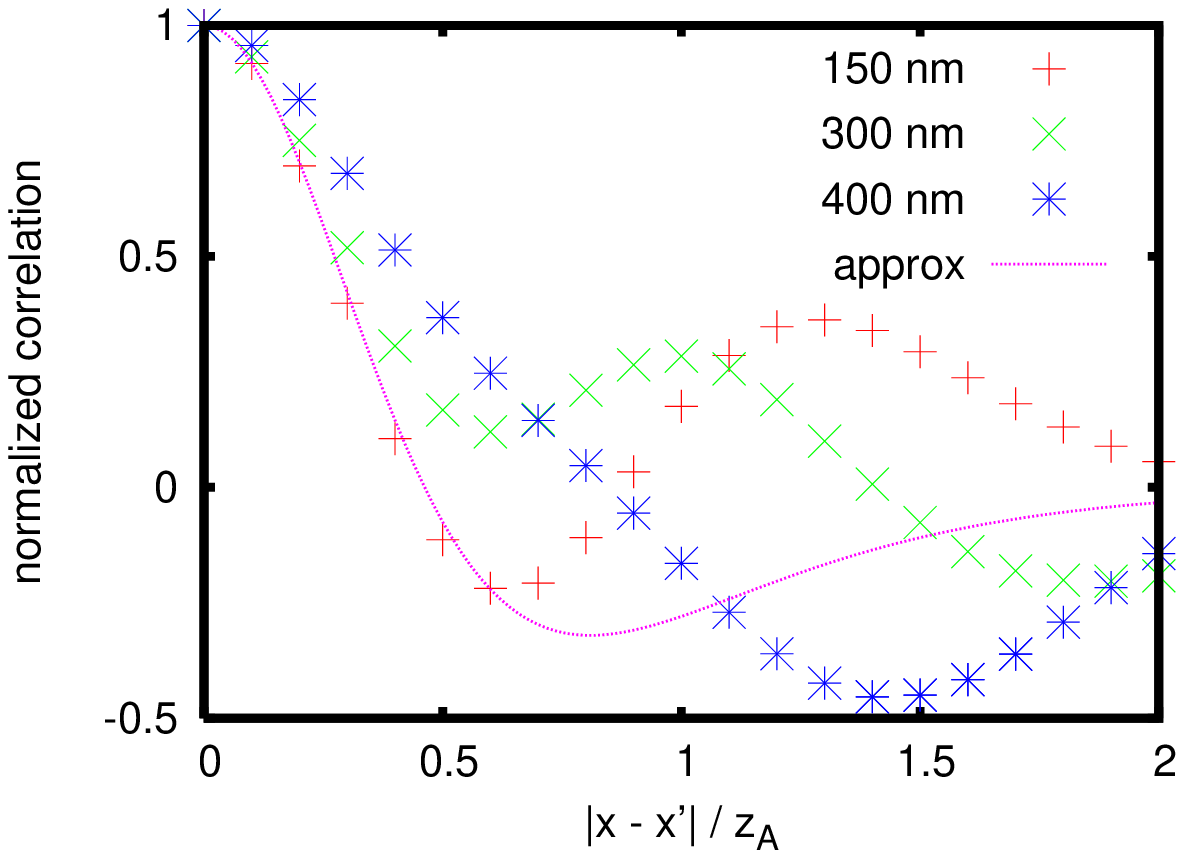, width=0.45\textwidth}
  \caption{\label{Fig:CorrelationNumeric} (Color online)
    Plot of the correlation functions (left) $\bigl\langle \Gamma^{(1)}_\perp (\mathbf{r}) \Gamma^{(1)}_\perp(\mathbf{r}') \bigr\rangle$ 
    and (right) $\bigl\langle \Gamma^{(1)}_\parallel (\mathbf{r}) \Gamma^{(1)}_\parallel(\mathbf{r}') \bigr\rangle$ 
    in Eq.~(\ref{Eq:CorrInText}) for different distances $z$ at $\omega_0 = 1.787\cdot10^{14}\,{\rm s}^{-1}$ choosing the 
    surface roughness parameters $\delta = 5\,{\rm nm}$ and $a = 50\,{\rm nm}$ (\ref{Eq:CorrelParLarge}). The correlation functions are normalized 
    to their value at $\mathbf{x} = \mathbf{x}'$. Furthermore we plot again the approximations shown in Fig.~\ref{Fig:Correlation}.
  }
\end{figure}

In Fig.~\ref{Fig:Correlation} we plot the quasistatic results of $\bigl\langle \Gamma^{(1)}_\perp (\mathbf{x}) \Gamma^{(1)}_\perp(\mathbf{x}') \bigr\rangle$ 
and $\bigl\langle \Gamma^{(1)}_\parallel (\mathbf{x}) \Gamma^{(1)}_\parallel(\mathbf{x}') \bigr\rangle$ from Eq.~(\ref{Eq:CorrelPerpLarge}) and (\ref{Eq:CorrelParLarge}) 
for a fixed distance $z_\rA$ which is assumed to be so small that the conditions for the quasistatic approximations are met, but $z_\rA \gg a$.
Note that the regime where these approximations are valid might be hard to achieve in practice, since the non-retarded regime starts for SiC for example 
for $z_\rA < 200\,{\rm nm}$, i.e., for distances which are not much larger than typical surface roughness correlation lengths. 
Nonetheless, one can expect that, for intermediate region, the correlation length is the atom-surface distance $z_\rA$. To illustrate this fact, we plot in 
Fig.~\ref{Fig:CorrelationNumeric} numerical results for the correlation function in Eq.~(\ref{Eq:CorrInText}). It can be seen that for intermediate 
distances $a < z_\rA < \lambda$ the correlation length is about $z_\rA$. 

\subsection{Variance and standard deviation}

\begin{figure}[Hhbt]
  \centering
  \epsfig{file=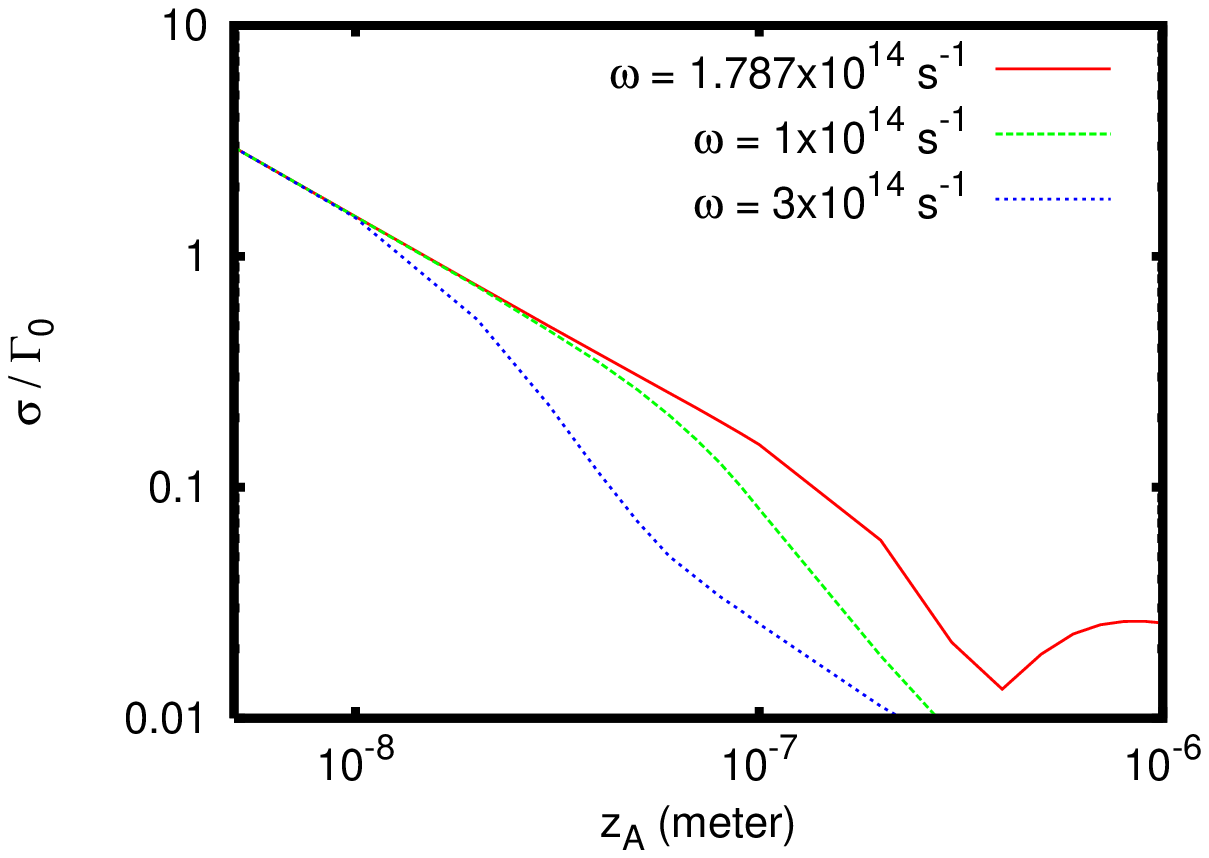, width=0.45\textwidth}
  \epsfig{file=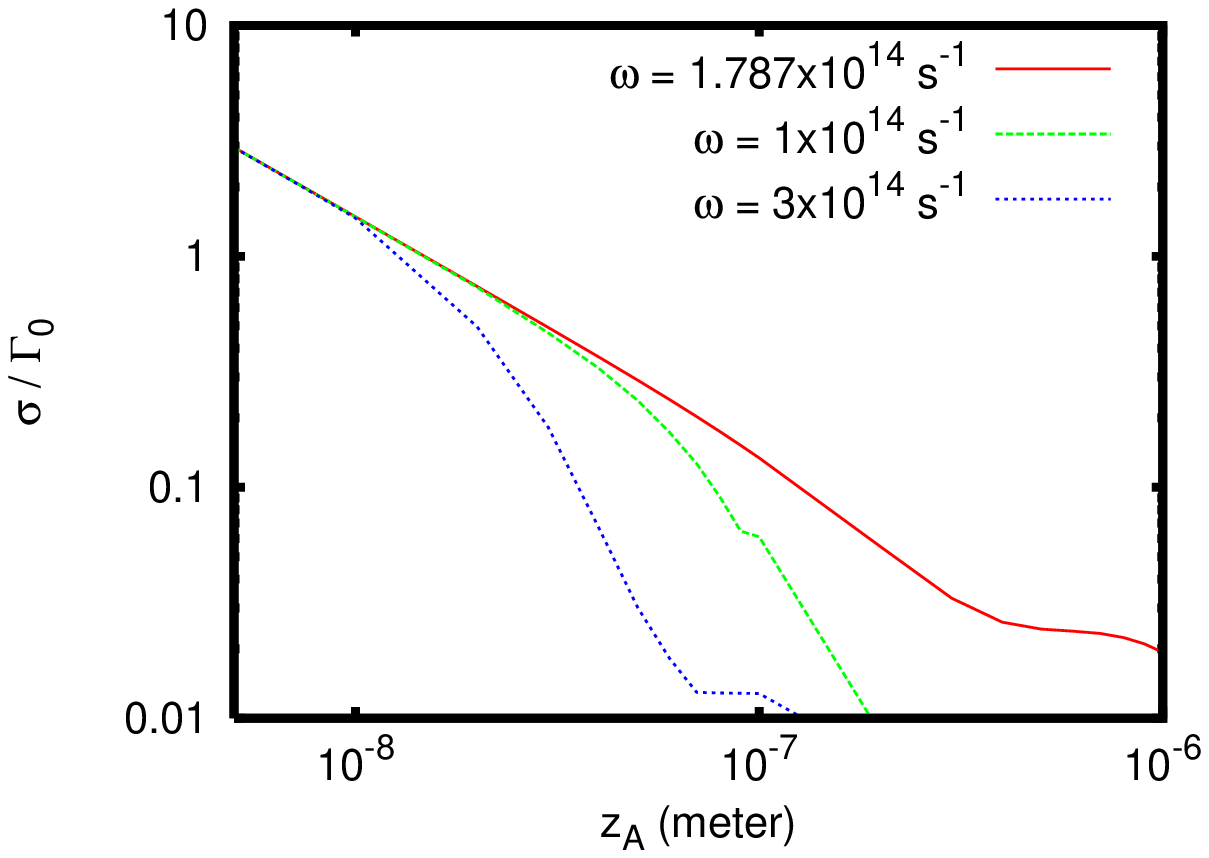, width=0.45\textwidth}
  \caption{\label{Fig:VarianzPerp}(Color online)
            Plot of the standard deviation $\sigma_\perp$ (left) and $\sigma_\parallel$ (right) for different distances of an atom placed at
           the distance $z_\rA$ above a rough SiC surface with $\delta = 5\,{\rm nm}$ and $a = 200\,{\rm nm}$. The transition frequencies are chosen to be
           $\omega_0 = 1\cdot10^{14}\,{\rm s}^{-1},  1.787\cdot10^{14}\,{\rm s}^{-1}  $ and $3\cdot10^{14}\,{\rm s}^{-1}$.
  }
\end{figure}

Let us now return to the variance given by Eq.~(\ref{Eq:Variance}). 
In the quasistatic limit using Eq.~(\ref{Eq:abiggerz}) we find for $z \ll a$ the simple expression
\begin{equation}
  \frac{\sigma^2_{\parallel/\perp}}{\biggl[\Gamma_{\parallel/\perp}^{(0)}(z)\biggr]^2 } \approx 9 \frac{\delta^2}{z^2_\rA}.
\label{Eq:PAVariance}
\end{equation}
In Fig.~\ref{Fig:VarianzPerp} we show some plots of the standard deviation $\sigma/\Gamma_0$. It can be seen that 
the standard deviation approaches the approximate result $3 \delta/z_\rA$ for $z_\rA \ll a$ so that 
for very small distances as $z = 10\,{\rm nm}$ the standard deviation or variance is on the order of 
$\Gamma^{(0)}$ or $[\Gamma^{(0)}]^2$, resp., 
indicating that fluctuations of the decay rate are large in the quasi-static regime. 
In all shown cases $\sigma$ falls off rapidly with the surface atom distance, which is due to the small 
roughness considered. It could be expected from Eq.~(\ref{Eq:CorrelPerp}) that in the distance regime $a \ll z_\rA$ the
standard deviation varies like $\sigma \propto \delta a /z^2_\rA$. Indeed, in Fig.~\ref{Fig:VarianzPerp} we find this power law 
for $\sigma_\perp$ but not for $\sigma_\parallel$. Note, that for the distance region around $200\,{\rm nm}$ for 
which we have found a relatively large roughness correction of $15\%$ to the mean decay rate  
$\sigma_{\perp/\parallel}$ is smaller than $10\%$.  

\subsection{LDOS fluctuations and $C_0$ correlation}

Finally we want to explore the relation between the LDOS and the infinite range intensity correlation $C_0$ as studied
for multiple scattering 
media~\cite{Shapiro1999}. 
It was shown that this infinite range correlation equals the LDOS fluctuations~\cite{VanTiggelen2006}. 
Since the decay rate is proportional to the LDOS, one could expect by analogy that above a rough surface 
the $C_0$ correlation equals the decay rate fluctuations or more precisely 
$C_0 = \sigma^2 / \langle \Gamma \rangle^2$. 
To prove this, we follow the reasoning of Ref.~\cite{CazeEtAl2010} adapted to our problem. First we assume for simplicity
that we have a non-absorbing half-space with a rough surface. Then the radiated power of a dipole in vacuum $P_0$
and of a dipole above the rough surface $P$ can be related to the decay rate of an atom in vacuum $\Gamma_0$
and the decay rate of an atom above a rough surface $\Gamma$ by the simple relation~\cite{NovotnyHecht2006}
\begin{equation}
  \frac{P}{P_0} = \frac{\Gamma}{\Gamma_0}.
\end{equation}
Since the decay rate is proportional to the LDOS $\rho$ at the position of the atom we 
also have $P/P_0 = \rho/\rho_0$. 

Now, we define the speckle correlation function $C(\mathbf{x,x'})$ for the lateral correlations with respect to $\mathbf{x}$ and
$\mathbf{x}'$ above the surface as
\begin{equation}
  C(\mathbf{x,x'}) = \frac{\langle I(\mathbf{x}) I(\mathbf{x}')\rangle}{\langle I(\mathbf{x})\rangle \langle I(\mathbf{x}')\rangle} - 1
\end{equation}
where $I(\mathbf{x})$ is the radiated power of the dipole at the position $\mathbf{x}$ such that the integral
over a plane parallel to the mean surface at $z = 0$ (but for $z > z_\rA$ and $z>\lambda_0$ so that evanescent waves are not included)
\begin{equation}
  \lim_{r_0 \rightarrow \infty} \frac{1}{A} \int_A \rd^2 x\, I(\mathbf{x}) = P \cdot R
\end{equation}
gives the total radiated power into the halfspace for $z > z_\rA$. Here, we have introduced the circular area $A = \pi r_0^2$. 
The factor $R$ takes into account that only a part of the total power $P$ is radiated into the halfspace for $z > z_\rA$, whereas another
part $P \cdot T$ is radiated into the halfspace $z < 0$ such that the total
power is the sum of both contributions, i.e., $T + R = 1$.  

With the relation $P/P_0 = \rho/\rho_0$ and the definition of $I(\mathbf{x})$ and $C(\mathbf{x,x'})$ we have
\begin{equation}
\begin{split}
  \biggl\langle \frac{\rho^2}{\rho_0^2} \biggr\rangle 
           &= \lim_{r_0 \rightarrow \infty} \frac{1}{A^2} \frac{1}{R^2} \frac{1}{P_0^2} \int_A \rd^2 x \int_A \rd^2 x'\, \langle I(\mathbf{x}) I(\mathbf{x}') \rangle \\
           &= \lim_{r_0 \rightarrow \infty} \frac{1}{A^2} \frac{1}{R^2} \frac{1}{P_0^2} \int_A \rd^2 x \int_A \rd^2 x'\, \langle I(\mathbf{x}) \rangle \langle I(\mathbf{x}') \rangle \bigl[ C(\mathbf{x,x'}) + 1 \bigr].          
\end{split}
\label{Eq:Rho}
\end{equation}
Since after averaging we retrieve the translational invariance parallel to the plane with $z = 0$ and isotropy, we
have $\langle I(\mathbf{x}) \rangle = P \cdot R = P_0 \cdot R \rho /\rho_0$. Inserting this relation into~(\ref{Eq:Rho})
it follows that
\begin{equation}
\begin{split}
  \langle \rho^2 \rangle &= \langle \rho \rangle^2 \lim_{r_0 \rightarrow \infty} \frac{1}{A^2} \int_A \rd^2 x \int_A \rd^2 x'\, \bigl[ 1 + C(|\mathbf{x - x'}|) \bigr] \\
                         &= \langle \rho \rangle^2 \biggl[1 + \lim_{r_0 \rightarrow \infty} \frac{2 \pi}{A} \int_0^\infty \rd r\, r C(r)\biggr]. 
\end{split}
\label{Eq:ResultA}
\end{equation}
In fact, the integral over $C(r)$ gives nonzero contributions only for $C(r) = {\rm const} \equiv C_0$, when assuming that 
for intensity correlations the relation $C(0) \geq C(r)$ for $r > 0$ is valid. 
Therefore the integral in Eq.~(\ref{Eq:ResultA}) reduces to $C_0$, i.e., 
the constant component of $C(r)$ which is the searched for infinite range $C_0$ correlation. Hence, we find
\begin{equation}
  C_0 = \frac{\langle \rho^2 \rangle - \langle \rho \rangle^2}{\langle \rho \rangle^2} = \frac{\langle \Gamma^2 \rangle - \langle \Gamma \rangle^2}{\langle \Gamma \rangle^2},
\end{equation}
which proves our statement that the $C_0$ correlation equals the decay rate fluctuations or $C_0 = \sigma^2 / \langle \Gamma \rangle^2$. 

Finally, using the perturbation expansion for the decay rate, we get for the $C_0$ correlations above a rough surface 
\begin{equation}
  C_0 = \frac{\sigma^2}{\langle \Gamma \rangle^2} \approx \frac{\langle \Gamma^{(1)} \Gamma^{(1)} \rangle}{\bigl(\Gamma^{(0)} + \langle \Gamma^{(2)} \rangle \bigr)^2} = \frac{\langle \Gamma^{(1)} \Gamma^{(1)} \rangle}{{\Gamma^{(0)}}^2} \frac{1}{\biggl(1 + \frac{\langle \Gamma^{(2)} \rangle}{\Gamma^{(0)}}\biggr)^2} \approx \frac{\langle \Gamma^{(1)} \Gamma^{(1)} \rangle}{{\Gamma^{(0)}}^2}.
\end{equation}
Hence, the square of the normalized standard deviation $\sigma/\Gamma^{(0)}$ gives the infinite range $C_0$ correlation
showing its sensitivity to the local environment which enters through its dependence on $z_\rA$. In the quasi-static
regime, we can now use the above derived result in Eq.~(\ref{Eq:PAVariance}) so that for $z_\rA \ll a$ we 
find  $C_0 \approx 9 \delta^2/z^2_\rA$. We note that this result is very similar to the result found by Shapiro~\cite{Shapiro1999}
for random media, where $C_0 = \pi/(k l)$ with the wave number $k$ and the mean free path of radiation inside the
random medium $l$. In our case, $\delta^2/z_\rA^2$ corresponds to the scattering strength $1/(k l)$.

%
%

\section{Conclusion}

We have studied the impact of surface roughness on the decay rate or inverse lifetime of a molecule or atom
above a rough surface. For pedagogical reasons we have only considered SiC as the bulk material, 
but the conclusions can be easily transfered to other dielectric materials supporting surface modes as for example silica. 
Our results show that the decay rate might be reduced by $15\%$ due to
the surface roughness for very shallow roughnesses with an rms of $\delta = 5\,{\rm nm}$
and a correlation length of $a = 200\,{\rm nm}$. This reduction is due to the surface induced
scattering of surface modes which prevails for intermediate distances. On the other hand, for
very small distances $z_\rA \ll a$ the rouhgness correction to the decay rate $\Gamma^{(0)}$ is 
due to the local electrostatic interaction of the atom or molecule with the surface
and is given by the simple expression $6 \delta^2/z_\rA^2 \Gamma^{(0)}$.   

In addition, we have studied the variance and the lateral correlation of decay rates above the rough surface.
We find that the lateral correlation length is approximately given by the distance $z_\rA$ itself in the nonretarded 
regime for distances larger than $a$. For distances smaller than $a$ the lateral correlation function resembles 
the surface roughness correlation function allowing for a direct measurement of the surface roughness properties by 
measuring decay rate or lifetime correlations. The variance itself is a special case of the lateral correlation function
and we have pointed out that it equals the $C_0$ correlation as for random media. We have shown that it can also 
be approximated by a simple result $\sigma^2 = 9 \delta^2/z_\rA^2$ in the quasistatic regime for $z_\rA \ll a$ 
emphasizing that the infinite range $C_0$ correlation highly depends on the local environment of the atom, i.e.,
on the distance $z_\rA$.

%
%

\begin{acknowledgements}
S.-A.\ B. gratefully acknowledges support from the Deutsche 
Akademie der Naturforscher Leopoldina (Grant No.\ LPDS 2009-7).
\end{acknowledgements}

%
%

\appendix

%
%

\section{Green's function for a flat surface}
\label{App:GreenFlat}

The Green's function with observation point and source point above the flat surface, i.e.,
for $0 < z \leq z'$ can be stated as~\cite{Sipe1987,HenkelSandoghdar1998}
\begin{equation}
\begin{split}
  \mathds{G}^{(0)}(\mathbf{r,r'};\omega) &= \int \!\!\!\frac{\rd^2 \kappa}{(2 \pi)^2} 
                                           \frac{\ri \re^{\ri \boldsymbol{\kappa\cdot(\mathbf{x - x'})}}}{2 \gamma_\rr} 
                                           \biggl[ \mathds{1}_{--} \re^{\ri \gamma_\rr (z' - z)} 
                                          + \mathds{R}_{+-} \re^{\ri \gamma_\rr (z' + z)} \biggr] \\
                                          &\quad - \frac{1}{3 k_0^2} \delta(z - z') \delta(\mathbf{x - x'}) \mathbf{e}_z \otimes \mathbf{e}_z  
\end{split}
\label{Eq:AppGreenFlat}
\end{equation} 
where $ \mathbf{e}_z$ is the unit vector in z-direction and $\otimes$ symbolizes the dyadic product. The
tensors $\mathds{1}$ and $\mathds{R}$ are defined as
\begin{align}
  \mathds{1}_{--} &= \sum_{j = \{\rs,\rp\}} \hat{\mathbf{a}}_j^-(\boldsymbol{\kappa}) \otimes \hat{\mathbf{a}}_j^-(\boldsymbol{\kappa}) \\
  \mathds{R}_{+-} &=  \sum_{j = \{\rs,\rp\}} r_j \hat{\mathbf{a}}_j^+(\boldsymbol{\kappa}) \otimes \hat{\mathbf{a}}_j^-(\boldsymbol{\kappa})
\end{align}
where 
\begin{align}
  \hat{\mathbf{a}}_\rs^-(\boldsymbol{\kappa}) &= \hat{\mathbf{a}}_\rs^+ (\boldsymbol{\kappa}) = \frac{1}{\kappa} (-k_y, k_x,0)^\rt , \\
  \hat{\mathbf{a}}_\rp^-(\boldsymbol{\kappa}) &= - \frac{1}{\kappa k_0} (k_x \gamma_\rr, k_y \gamma_\rr, \kappa^2)^\rt , \\
  \hat{\mathbf{a}}_\rp^+(\boldsymbol{\kappa}) &= \frac{1}{\kappa k_0} (k_x \gamma_\rr, k_y \gamma_\rr, -\kappa^2)^\rt
\end{align}
are the polarization vectors for s- and p-polarization. Note that these vectors are always orthogonal, but only normalized
for propagating modes with $\kappa < k_0$. The reflection coefficients $r_\rs$ and $r_\rp$ are the usual Fresnel coefficients
\begin{equation}
  r_\rs = \frac{\gamma_\rr - \gamma_\rt}{\gamma_\rr + \gamma_\rt} \quad\text{and}\quad 
  r_\rp = \frac{\gamma_\rr \epsilon - \gamma_\rt}{ \gamma_\rr \epsilon + \gamma_\rt}.
\end{equation}

\section{First-order Green's function}

The correction to the Green's function we find from first-order perturbation theory is~\cite{Greffet1988}
\begin{equation}
  \mathds{G}^{(1)}(\mathbf{r,r'};\omega) = - \int\!\!\!\frac{\rd^2 \kappa}{(2 \pi)^2} \int\!\!\!\frac{\rd^2 \kappa'}{(2 \pi)^2}
                                           \frac{k_0^2 (\epsilon - 1)}{4 \gamma_\rr \gamma_\rr'} \re^{\ri (\boldsymbol{\kappa}\cdot\mathbf{x} + \gamma_\rr z)}
                                           \tilde{S}^{(1)} (\boldsymbol{\kappa}' - \boldsymbol{\kappa}) \re^{\ri (\boldsymbol{\kappa}'\cdot\mathbf{x}' + \gamma_\rr' z')} \mathds{X}_{+-}(\boldsymbol{\kappa},\boldsymbol{\kappa}') 
\label{Eq:FirstOrderGreensFunction}
\end{equation}  
with
\begin{equation}
   \tilde{S}^{(1)} (\boldsymbol{\kappa}' - \boldsymbol{\kappa}) = \int\!\!\rd^2 x \, \re^{-i(\boldsymbol{\kappa}' - \boldsymbol{\kappa}) \cdot \mathbf{x}}S(\mathbf{x}).
\end{equation}
and
\begin{equation}
  \mathds{X}_{+-}(\boldsymbol{\kappa},\boldsymbol{\kappa}') = \sum_{i,j = \{\rs,\rp\}} \hat{\mathbf{a}}^+_i(\boldsymbol{\kappa}) \otimes \hat{\mathbf{a}}^-_j(\boldsymbol{\kappa}') X_{ij} (\boldsymbol{\kappa},\boldsymbol{\kappa}').
\end{equation}
The elements of the tensor $X_{ij}$ are given as
\begin{align}
  X_{\rs\rs} &= t_\rs t_\rs' \hat{\boldsymbol{\kappa}}\cdot \hat{\boldsymbol{\kappa}}' \\
  X_{\rs\rp} &= - t_\rs t_\rp' \frac{\gamma_\rt'}{\sqrt{\epsilon} k_0} \mathbf{e}_z \cdot \hat{\boldsymbol{\kappa}}\times\hat{\boldsymbol{\kappa}}' \\
  X_{\rp\rs} &= - t_\rp t_\rs' \frac{\gamma_\rt}{\sqrt{\epsilon} k_0} \mathbf{e}_z \cdot \hat{\boldsymbol{\kappa}}\times\hat{\boldsymbol{\kappa}}' \\
  X_{\rp\rp} &= + t_\rp t_\rp' \frac{1}{\epsilon k_0^2} (\kappa \kappa' \epsilon - \gamma_\rt \gamma_\rt' \hat{\boldsymbol{\kappa}}\cdot \hat{\boldsymbol{\kappa}}' )
\end{align}
where $\hat{\boldsymbol{\kappa}} =\boldsymbol{\kappa}/\kappa$ and $t_\rs, t_\rp$ are the usual amplitude transmission coefficients
\begin{equation}
  t_\rs = \frac{2 \gamma_\rr}{\gamma_\rr + \gamma_\rt} \qquad\text{and}\qquad 
   t_\rp = \frac{2 \gamma_\rr \sqrt{\epsilon}}{\gamma_\rr \epsilon + \gamma_\rt}.
\end{equation}

%
%

\section{Correlation function}
\label{App:Correlation}

By inserting the Green's function from Eq.~(\ref{Eq:FirstOrderGreensFunction}) into Eq.~(\ref{Eq:PurcellFactor}) we find for
the first-order correction to the decay rate the expression
\begin{equation}
  \frac{\Gamma^{(1)}_i(\mathbf{r})}{\Gamma_\infty} = \frac{3 \pi \ri}{k_0} \int\!\!\frac{\rd^2 \kappa}{(2 \pi)^2} \int\!\!\frac{\rd^2 \kappa'}{(2 \pi)^2}
                   \bigl[ \tilde{S}^{(1)} (\boldsymbol{\kappa}' - \boldsymbol{\kappa}) a_i (\boldsymbol{\kappa},\boldsymbol{\kappa}';\mathbf{r}) \\
                         - \tilde{S}^{(1)} (\boldsymbol{\kappa} - \boldsymbol{\kappa}') a_i^* (\boldsymbol{\kappa},\boldsymbol{\kappa}';\mathbf{r})\bigr]\re^{\ri \mathbf{x} \cdot (\boldsymbol{\kappa} - \boldsymbol{\kappa}')} 
\end{equation}
where
\begin{equation}
  a_i (\boldsymbol{\kappa},\boldsymbol{\kappa}';\mathbf{r}) = \frac{k_0^2 (\epsilon - 1)}{4 \gamma_\rr \gamma_\rr'} \re^{\ri z (\gamma_\rr + \gamma_\rr')} \bigl[ \mathbf{e}_i^\rt \cdot \mathds{X}_{+-} \cdot \mathbf{e}_i \bigr].
\label{Eq:int_a}
\end{equation}
With this definition at hand it is an easy task to check that the correlation function is
\begin{equation}
\begin{split}
 \frac{\bigl\langle \Gamma^{(1)}_i(\mathbf{r}) \Gamma^{(1)}_j(\mathbf{r}') \bigr\rangle}{\Gamma_\infty^2} &= \frac{(3 \pi)^2}{k_0^2} 2 \Re \int\!\!\frac{\rd^2 \kappa}{(2 \pi)^2} \int\!\!\frac{\rd^2 \kappa'}{(2 \pi)^2} \delta^2 g(|\boldsymbol{\kappa} - \boldsymbol{\kappa}'|) a_i (\boldsymbol{\kappa},\boldsymbol{\kappa}';\mathbf{r}) \\
                            &\quad\times  \int\!\!\frac{\rd^2 \kappa''}{(2 \pi)^2} [ a_j^* (\boldsymbol{\kappa}'',\boldsymbol{\kappa}^-;\mathbf{r}')
                            - a_j (\boldsymbol{\kappa}'',\boldsymbol{\kappa}^+;\mathbf{r}') ]\re^{\ri (\mathbf{x} - \mathbf{x}') \cdot (\boldsymbol{\kappa} - \boldsymbol{\kappa}')} 
\end{split}
\label{Eq:CorrFuncApp}
\end{equation}
using the relations
\begin{align}
  \boldsymbol{\kappa}^+ = \boldsymbol{\kappa}'' + (\boldsymbol{\kappa} - \boldsymbol{\kappa}'), \\
  \boldsymbol{\kappa}^- = \boldsymbol{\kappa}'' - (\boldsymbol{\kappa} - \boldsymbol{\kappa}').  
\end{align}
By introducing the new variable $\boldsymbol{\xi} = \boldsymbol{\kappa} - \boldsymbol{\kappa}'$ we can write the correlation function
as
\begin{equation}
  \frac{\bigl\langle \Gamma^{(1)}_i(\mathbf{r}) \Gamma^{(1)}_j(\mathbf{r}') \bigr\rangle}{\Gamma_\infty^2} = \frac{(3 \pi)^2}{k_0^2} 2 \Re \int\!\!\frac{\rd^2 \xi}{(2 \pi)^2} \delta^2 g(|\boldsymbol{\xi}|) 
                        F_j (\boldsymbol{\xi};z') G_i (\boldsymbol{\xi};z) \re^{\ri \boldsymbol{\xi}\cdot(\mathbf{x} - \mathbf{x}')}
\label{Eq:CorrRef}
\end{equation}
with
\begin{equation}
   G_i (\boldsymbol{\xi};z) =  \int\!\!\frac{\rd^2 \kappa}{(2 \pi)^2} a_i (\boldsymbol{\kappa},\boldsymbol{\kappa} - \boldsymbol{\xi};z)
\label{Eq:G}
\end{equation}
and
\begin{equation}
  F_j (\boldsymbol{\xi};z') = G_j^* (\boldsymbol{\xi};z') - G_j (-\boldsymbol{\xi};z').
\label{Eq:F}
\end{equation}

%
%

\section{Approximations for quasi-static limit}
\label{App:QuasiStaticAppr}

In the quasi-static limit ($\kappa \gg k_0$) the reflection coefficients can be approximated by
\begin{equation}
  r_\rp \approx \frac{\epsilon - 1}{\epsilon + 1} \quad\text{and}\quad r_\rs \approx \frac{\epsilon - 1}{4} \frac{k_0^2}{\kappa^2}.
\end{equation}
By inserting these relations into Eqs.~(\ref{Eq:G}) we find the quasi-static approximations for
the decay rates
\begin{align}
  \frac{\Gamma^{(0)}_\parallel}{\Gamma_\infty} &\approx \frac{3}{16} \frac{1}{(k_0 z)^3} \Im \biggl( \frac{\epsilon - 1}{\epsilon + 1} \biggr) ,\\
  \frac{\Gamma^{(0)}_\perp}{\Gamma_\infty}     &\approx \frac{3}{8} \frac{1}{(k_0 z)^3} \Im \biggl( \frac{\epsilon - 1}{\epsilon + 1} \biggr) .
\end{align}
In particular, $\Gamma_\perp^{(0)} = 2 \Gamma_\parallel^{(0)}$.

\subsection*{i) distance regime $z \ll a$}

Now, we want to find similar simple approximate expressions for the correlation function in Eq.~(\ref{Eq:CorrRef}). To this end
we consider first $\kappa \gg \xi$, which is fulfilled for $z \ll a$. For such wave vectors we can approximate Eq.~(\ref{Eq:G})
by 
\begin{align}
  G_\parallel (\boldsymbol{\xi}; z) &\approx \int \!\! \frac{\rd \kappa}{2 \pi}\, \frac{k_0^2(\epsilon - 1)}{8 \gamma_\rr^2} \re^{2 \ri \gamma_\rr z} 
                             \biggl( t_\rs^2 - \frac{t_\rp^2}{\epsilon k_0^2} \frac{\gamma_\rr^2}{k_0^2} \bigl[ \kappa^2 (\epsilon + 1) - k_0^2 \epsilon \bigr] \biggr), \\
  G_\perp (\boldsymbol{\xi}; z) &\approx \int \!\! \frac{\rd \kappa}{2 \pi}\, \frac{k_0^2(\epsilon - 1)}{4 \gamma_\rr^2} \re^{2 \ri \gamma_\rr z} 
                            \frac{t_\rp^2}{\epsilon k_0^2} \frac{-\kappa^2}{k_0^2} \bigl[ \kappa^2 (\epsilon + 1) - k_0^2 \epsilon \bigr].
\end{align}
Using the quasi-static approximation for the transmission coefficients 
\begin{equation}
  t_\rs \approx 1 \quad\text{and} \quad t_\rp \approx \frac{2 \sqrt{\epsilon}}{\epsilon + 1}
\end{equation}
allows for further simplification. We find
\begin{align}
   G_\parallel (\boldsymbol{\xi}; z) &\approx - \frac{3!}{4 \pi} \frac{1}{(2 z)^4} \frac{1}{k_0^2} \frac{\epsilon - 1}{\epsilon + 1}, \\
   G_\perp (\boldsymbol{\xi}; z) &\approx - \frac{ 3!}{2 \pi} \frac{1}{(2 z)^4} \frac{1}{k_0^2} \frac{\epsilon - 1}{\epsilon + 1} . 
\end{align}
Inserting these approximations into Eq.~(\ref{Eq:CorrRef}) finally yields
\begin{align}
  \frac{\bigl\langle \Gamma^{(1)}_\parallel (\mathbf{r}) \Gamma^{(1)}_\parallel (\mathbf{r}') \bigr\rangle}{\Gamma_\infty^2} 
          &\approx \frac{81}{256} \frac{\delta^2}{k_0^6 z^4 {z'}^4}\biggl[\Im\biggl( \frac{\epsilon - 1}{\epsilon + 1} \biggr)\biggr]^2 W(|\mathbf{x - x'}|) ,  \\
   \frac{\bigl\langle \Gamma^{(1)}_\perp (\mathbf{r}) \Gamma^{(1)}_\perp (\mathbf{r}') \bigr\rangle}{\Gamma_\infty^2} 
          &\approx \frac{81}{64} \frac{\delta^2}{k_0^6 z^4 {z'}^4}\biggl[\Im\biggl( \frac{\epsilon - 1}{\epsilon + 1} \biggr)\biggr]^2 W(|\mathbf{x - x'}|) .
\end{align}
As can be expected from $G_\perp = 2 G_\parallel$ we find
\begin{equation}
  \frac{\bigl\langle \Gamma^{(1)}_\perp (\mathbf{r}) \Gamma^{(1)}_\perp (\mathbf{r}') \bigr\rangle }{\bigl\langle \Gamma^{(1)}_\parallel (\mathbf{r}) \Gamma^{(1)}_\parallel (\mathbf{r}') \bigr\rangle} = 4.
\end{equation}

\subsection*{ii) distance regime $z \gg a$}

In this limit, we consider the case $\kappa \gg \xi$ yielding 
\begin{align}
   G_\perp &\approx \int \frac{\rd^2 \kappa}{(2 \pi)^2} a_z (\boldsymbol{\kappa},-\boldsymbol{\xi}), \\
   G_\parallel &\approx \int \frac{\rd^2 \kappa}{(2 \pi)^2} a_x (\boldsymbol{\kappa},-\boldsymbol{\xi}). \\
\end{align}
Together with the quasi-static approximation, i.e, $\xi \gg k_0$ and $\xi \gg k_0 \epsilon$ we get
\begin{equation}
   G_\perp \approx \re^{-\xi z} \xi \int\frac{\rd \kappa}{2 \pi} \kappa^3 \frac{t_\rp (\epsilon - 1)}{k_0^2 4 \ri \gamma_\rr} \re^{\ri \gamma_\rr z} \frac{2 \sqrt{\epsilon}}{\epsilon + 1} 
           \equiv f_\perp (z)  \re^{-\xi z} \xi 
\end{equation}
and
\begin{equation}
   G_\parallel \approx \re^{-\xi z} \xi \cos^2(\theta) \frac{1}{2} \int\frac{\rd \kappa}{2 \pi} \kappa \frac{k_0^2 (\epsilon - 1)}{4 \ri \gamma_\rr} \re^{\ri \gamma_\rr z} \biggl[ 
                          \frac{\gamma_\rr \gamma_\rt t_\rp}{k_0^2 \epsilon} - \frac{r_\rs}{k_0^2 \sqrt{\epsilon}}\biggr] 
               \equiv  f_\parallel (z) \re^{-\xi z} \xi \cos^2(\theta)
\end{equation}
where we have introduced $\xi_x = \xi \cos{\theta}$ and $\xi_y = \xi \sin(\theta)$.
Finally, when plugging these results into Eq.~(\ref{Eq:CorrFuncApp}) we find~\cite{GradshteynRyzhik2007}
\begin{align}
   \frac{\bigl\langle  \Gamma^{(1)}_\perp (\mathbf{x})  \Gamma^{(1)}_\perp (\mathbf{x}') \bigr\rangle }{\Gamma_\infty^2} 
                        &\propto \frac{a^2 \delta^2}{[z^2 + |\mathbf{x} - \mathbf{x}'|^2]^2} 
                        P_3 \biggl( \frac{z}{\sqrt{z^2 + (\mathbf{x} - \mathbf{x}')^2}} \biggr), \label{Eq:CorrelPerpLarge}\\
   \frac{\bigl\langle  \Gamma^{(1)}_\parallel (\mathbf{x})  \Gamma^{(1)}_\parallel (\mathbf{x}') \bigr\rangle }{\Gamma_\infty^2} 
                        &\propto \frac{\delta^2 a^2}{z^4} \biggl[ \, \tensor[_2]{F}{_1} \biggl(2, \frac{5}{2};3; -\frac{|\mathbf{x}-\mathbf{x}'|^2}{z^2}\biggr) \nonumber \\
                        &\qquad - \frac{20}{3}\, \frac{|\mathbf{x} - \mathbf{x}'|^2}{z^2} \tensor[_2]{F}{_1}\biggl(3, \frac{7}{2};3; -\frac{|\mathbf{x}-\mathbf{x}'|^2}{z^2}\biggr) \biggr] \label{Eq:CorrelParLarge}
\end{align}
where $P_3$ is the Legendre polynomial of third power and $\tensor[_2]{F}{_1}$ is the hypergeometric function.

%
%
%

%
%

%
%

\end{document}